\documentclass[10pt,
			   aps,
			   prb,
			   twocolumn, 
			   nofootinbib, 
			   showkeys,           % show keywords
			   superscriptaddress  % for auto numbered affils
			   ]{revtex4-1}

\usepackage{amssymb,amsmath,bm}
\usepackage{cmap}
\usepackage{graphicx}
\usepackage{color}
\usepackage{hyperref}
\usepackage{listings}
\usepackage{booktabs}
\usepackage{multirow}
\usepackage{comment}
\hypersetup{
    unicode=false,           % non-Latin characters in Acrobat’s bookmarks
    pdftoolbar=true,         % show Acrobat’s toolbar?
    pdfmenubar=true,         % show Acrobat’s menu?
    pdffitwindow=false,      % window fit to page when opened
    pdfstartview={FitH},     % fits the width of the page to the window
    pdfauthor={Denis Sakhno}, % author
    colorlinks=true,         % false: boxed links; true: colored links
    linkcolor=blue,          % color of internal links
    citecolor=blue,           % color of links to bibliography
}
\usepackage{dsfont}
\usepackage{physics}

\usepackage[english]{babel}

\graphicspath{{fig/}}

\begin{document}

\title{Controlling the dispersion of longitudinal waves via the affine deformation of the interlaced wire medium}

\newcommand{\affilITMO}{ITMO University, Kronverksky pr. 49, 197101 St. Petersburg, Russia}

\author{Denis Sakhno}
\affiliation{\affilITMO}
\email{denis.sakhno@metalab.ifmo.ru}
\author{Eugene Koreshin}
\affiliation{\affilITMO}
\author{Pavel A. Belov}

\affiliation{\affilITMO}

%________________________________________Abstract
\begin{abstract}
We studied the dispersion properties of double interlaced wire metamaterials with geometry modified by affine transformation.
That metamaterials were found to support eigenmodes with longitudinal polarization at low frequencies for all deformations. Due to the spatial dispersion the metamaterials isofrequency surfaces are centered at the Brillouin zone edges (rather than at $\Gamma$-point) and have the shape of ellipsoids. The refractive indices corresponding to the ellipsoids were analyzed both analytically and numerically.
\end{abstract}
%________________________________________

\date{\today}
\maketitle

%________________________________________Introduction
\section{\label{sec:intro}Introduction}
Metamaterials are artificially engineered media with electromagnetic properties that are not typical for natural materials.
% The metanmaterials have been studied for the last decades and so far this topic has not exhausted itself.
Metamaterials made of wires, also referred to as wire media \cite{simovski2012wire} are one of the simplest (in terms of geometry) members of the class, on the other hand, they feature very sophisticated electromagnetic properties due to their nonlocal response. A two-dimensional single-periodic array of parallel wires, the so-called simple wire medium \cite{belov2003strong, SimovskiBelov2004}, transmits near-field distribution images to long distances with subwavelength resolution \cite{belov2008transmission}, and it is also used in magnetic resonance imaging (MRI) to improve image quality \cite{slobozhanyuk2016mri}. The non-connected double wire medium \cite{SimovskiBelov2004,silveirinha2005homo} exhibits hyperbolic behavior \cite{morgado2014reversed}, while the triple one has the specific property of having four optical axes \cite{sakhno2022quadraxial}, which once again emphasizes the dominant role of the non-local behavior of the materials discussed.

An interesting wire metamaterial is a three-dimensional metal mesh \cite{silveirinha2005homo, silveirinha2009artificial} that exhibits plasma-like behavior, i.e. prohibits the propagation of electromagnetic waves below a certain \textit{plasma frequency}, and also supports longitudinal modes above the latter. A large subclass of complex interlaced wire metamaterials \cite{fan2007nonmax} has been proposed based on the artificial plasma. They consist of several independent metallic lattices inserted one into the other.  Surprisingly, such materials were found to have $N-1$ low-frequency modes, where $N$ is the number of independent meshes.

The properties of double interlaced wire metamaterial dispersion have been studied, with results presented in \cite{fan2010transmission, silveirinha2017lighttun, powell2021dark, sakhno2021longitudinal}. Thus, the longitudinal nature of the only low-frequency mode was established \cite{silveirinha2017lighttun,sakhno2021longitudinal}, and it isofrequency contours  were shown to arise at the corners of the Brillouin zone \cite{sakhno2021longitudinal} for one of the material configurations (\textit{symmetrical}) composed of two identical cubic meshes.

The present article deals with the same double interlaced wire medium in its so-called symmetrical (body centered cubic) configuration as that discussed in  \cite{powell2021dark,sakhno2021longitudinal}. Using this metamaterial as the initial one and deforming it in a special way would make it possible to describe a new family of materials, the dispersion properties of which will be closely related to the chosen deformation. We will show that in the case of the double interlaced wire medium, the deformation will not affect the longitudinal polarization of the low-frequency mode and its \textit{splitting}, which confirms the scalar nature of the mode predicted in \cite{fan2007nonmax}.
%________________________________________

%________________________________________Object of study
\section{\label{sec:object} Object of study}
The object of the present research is the interlaced wire metamaterial \cite{fan2007nonmax, silveirinha2017lighttun, fan2010transmission, simovski2012wire}, which consists of two metal lattices inserted one into the other. In the \textit{original configuration} of the material, each of the identical lattices is cubic with the period $a$, and the metamaterial itself has body-centered cubic symmetry, the primitive unit cell of which is a rhombohedron (this is discussed in more detail in our previous work \cite{sakhno2021longitudinal}).

%________________________________________Fig 1
\begin{figure}[h!]
    \begin{minipage}[h]{1\linewidth}
	\center{\includegraphics[width=1\textwidth]{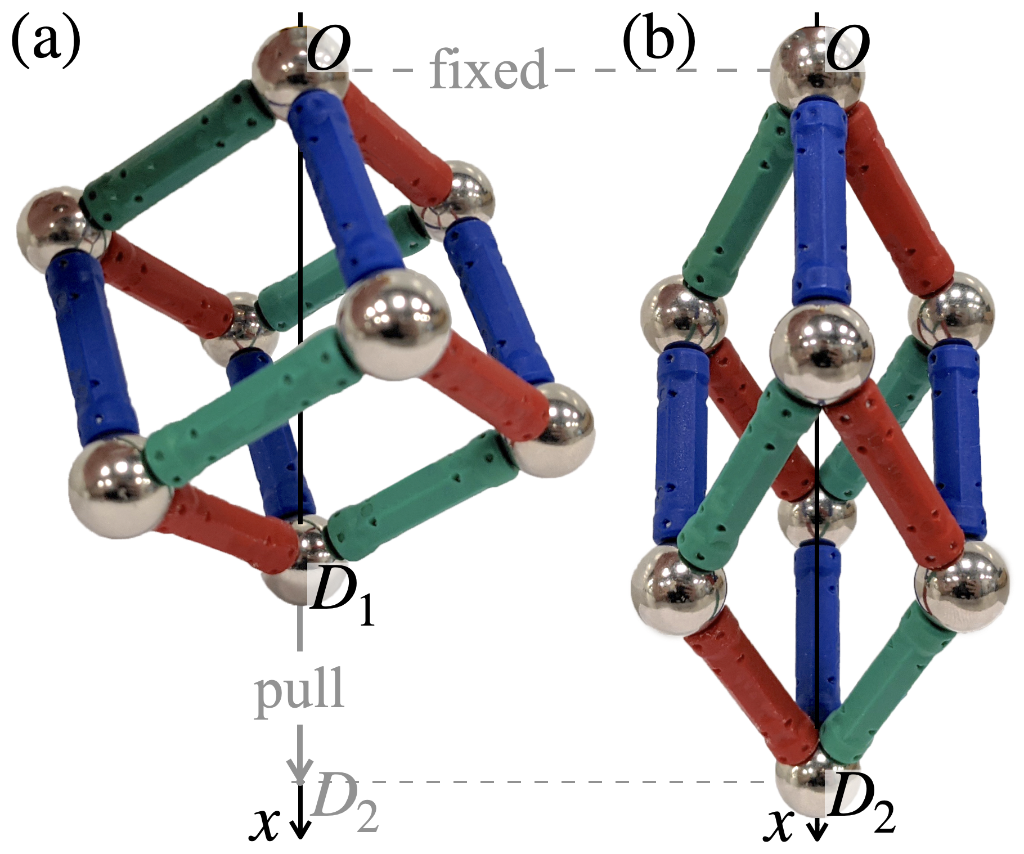}}
    \end{minipage}
    \caption{ (a) Cube skeleton from a magnetic kit for kids. (b) Trigonal trapezohedron sceleton resulting from stretching of the cube one.}
	\label{fig:geomag}
\end{figure}
%________________________________________

This paper addresses the issue of what will happen if the original lattices are deformed. For substantiation and for a clearer description of the deformation, we will use simple visualization.

First, we assemble a cube skeleton from a common magnetic kit for kids (Fig. \ref{fig:geomag}(a)), this cube will become an elementary block of one of the identical sublattices (in the original configuration of the material). The second step is to try and pull (or squeeze) it by two opposite ball-vertices (along one of the main diagonals), the skeleton will easily change its shape (Fig. \ref{fig:geomag}(b)). What is important about such deformation is that the lengths of the frame ribs remain unchanged, i.e. each face of the frame will be a rhombus. Regardless of the degree of stretching (or compression) the frame itself will retain the \textit{trigonal trapezohedron} form, since the main diagonal (along which the skeleton is stretched or compressed) will remain the 3-fold rotation axis for the skeleton and since all ribs are equal.

\begin{figure}[h!]
    \begin{minipage}[h]{1\linewidth}
        \center{\includegraphics[width=1\textwidth]{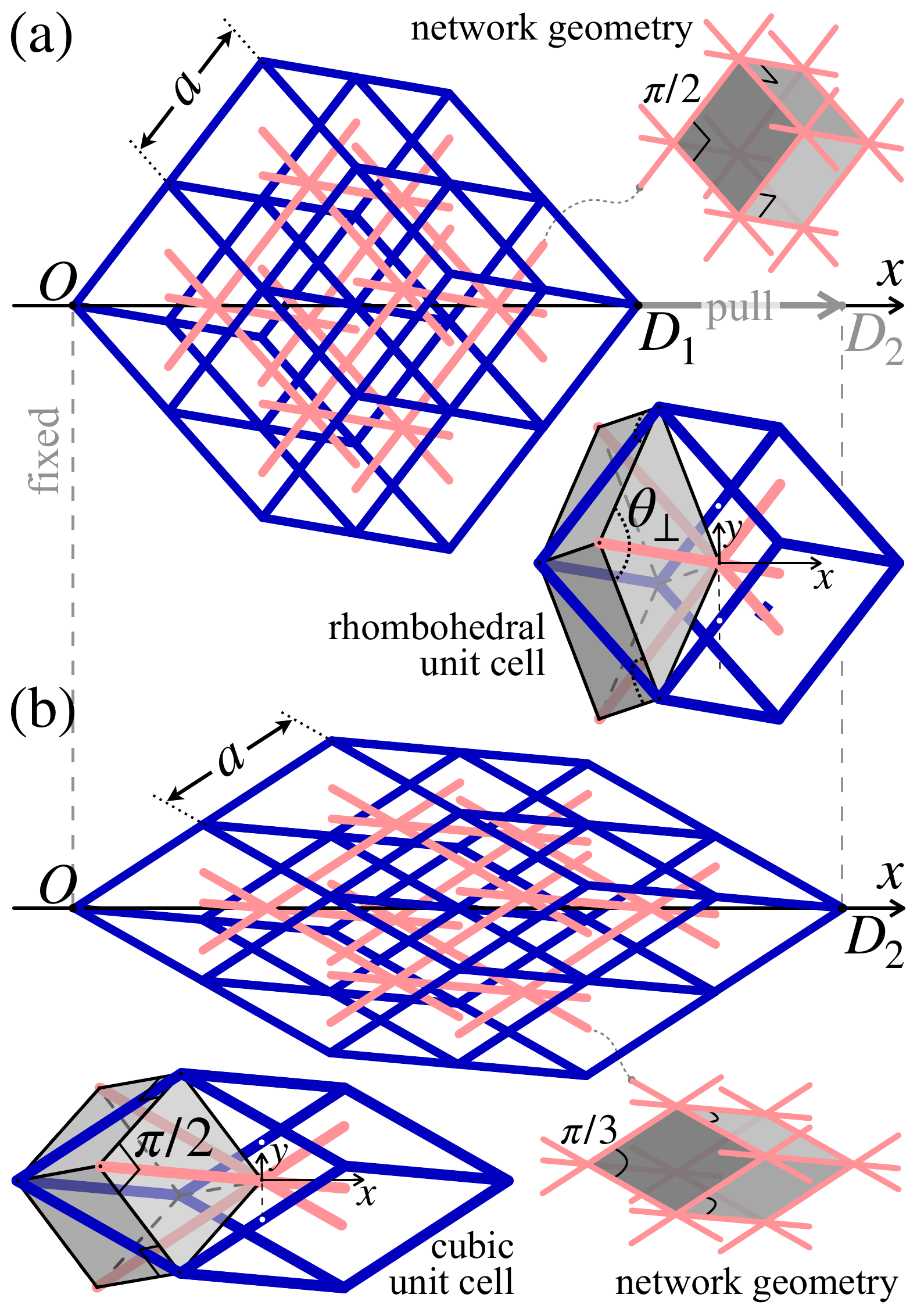}}
    \end{minipage}
    \caption{Double interlaced wire metamaterial consisting of two metal non-connected meshes (dark-blue and pink): (a) the \textit{original configuration} of the double interlaced wire metamaterial with a rhombohedral unit cell (composed of meshes with cubic cells, Fig. \ref{fig:geomag}(a)); (b) a metamaterial stretched along $x$-axis with a cubic unit cell (composed of meshes with rhombohedron cells, Fig. \ref{fig:geomag}(b)). Unit cells are shown in figure inserts.}
    \label{fig:struct}
\end{figure}

 It is important to mention that the lengths of the skeleton ribs (Fig. \ref{fig:geomag}) remain unchanged during our deformation. When the transformation is applied to the metamaterial, the wires length does not change, which means that the wire length $a$ is a \textit{deformation invariant} (Fig. \ref{fig:struct}).

%________________________________________Fig 2 (CHECK!)
%________________________________________

By applying this transformation to the entire metamaterial, i.e. deforming both lattices at the same time (keeping the half-period relative shift between them) we get a new \textit{family of materials}, each member of which is determined by the degree of stretching (or compression) of the original geometry (Fig. \ref{fig:struct}(a)). Further in this paper, we will refer to such new configurations as to a \textit{deformed} interlaced wire metamaterial, indicating the \textit{degree of deformation} (a corresponding parameter will be described later in Eq. (\ref{eq:p_param})).
%________________________________________

%________________________________________Deformation description
\section{\label{sec:geometry} Deformation  description}
Next, we describe the proposed deformation analytically by introducing the deformation operator. This operator will be applied to the interlaced wire metamaterial, i.e. to its elementary cell (unit cell of the \textit{original metamaterial} shown in Fig. \ref{fig:struct}(a)), which will retain the shape of a \textit{trigonal trapezohedron} (all its faces are identical rhombuses) like the skeleton assembled from the kit in Fig. \ref{fig:geomag}. Thus, any rhombohedron with the wires (of length $a$) in the equal faces diagonals will be an elementary cell of the metamaterial belonging to the \textit{family}. The simplest (in form) unit cell to imagine is the cubic cell shown in Fig. \ref{fig:struct}(b). What follows is the analytical description of the deformation operator. 

\begin{figure*}
    \begin{minipage}{1\linewidth}
        \center{\includegraphics[width=1\textwidth]{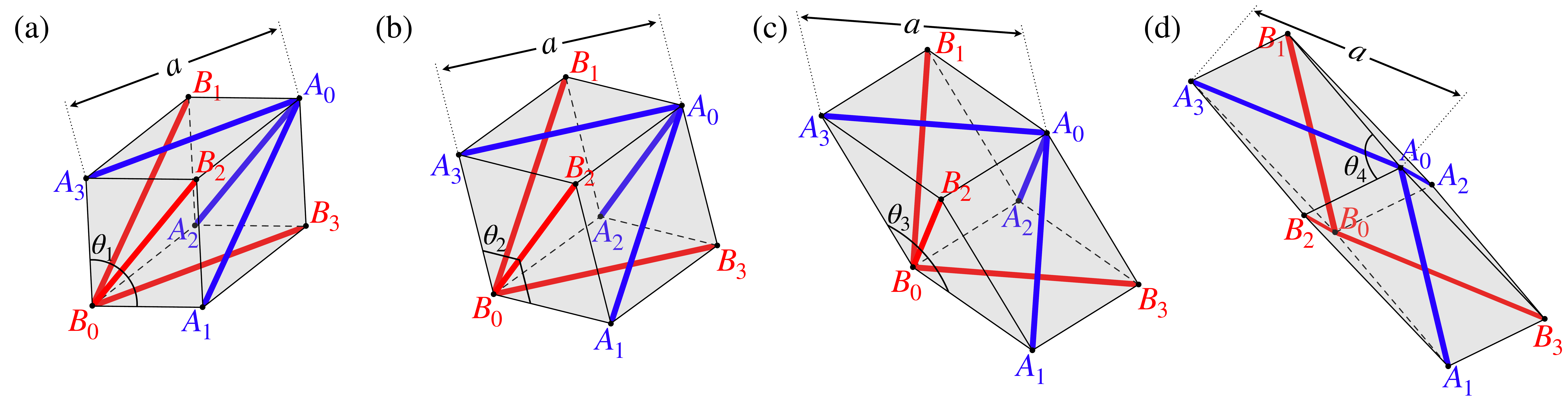}}
    \end{minipage}
    \caption{\textit{Trigonal trapezohedron} unit cells of metamaterials from \textit{the family}, whose geometry is defined by the \textit{deformation angle} (a) $\theta_1=5\pi/12$; (b) $\theta_2 = \pi/2$ (the cubic unit cell); (c) $\theta_3=\theta_\perp=\acos\left(-1/3\right)>\pi/2$ (the \textit{original} interlaced wire metamaterial); (d) $\theta_4=2\pi/3-\pi/90>\theta_\perp$. The length of all wires (blue or red lines) is the same in all cases ($a=A_0A_1=A_0A_2=A_0A_3=B_0B_1=B_0B_2=B_0B_3$). Axes: $OX\parallel B_0A_0$ and $OY\parallel B_1B_2\parallel A_1A_2$.}
    \label{fig:unitcell}
\end{figure*}

As mentioned above, the proposed deformation keeps the length of wires $a$ (i.e. periods of lattices) constant and preserves the 3-fold rotation axis $A_0B_0$ (see Fig. \ref{fig:unitcell}), which is the same as the axis of stretching (or compression). 

Any \textit{trigonal trapezohedron} and, thus, the \textit{modified} metamaterial unit cell can be restored using only two parameters (see Fig. \ref{fig:unitcell}(b)): (i) the face angle $\angle A_3B_0A_1=\theta$ ($\angle A_3B_0A_1=\angle A_3B_0A_2=\angle A_1B_0A_2=\angle B_1A_0B_2=\angle B_3A_0B_2=\angle B_3A_0B_2=\theta$), whose value is limited by the range $\theta \in \left(0, 2\pi / 3 \right)$ (the upper $\theta$-limit is $2\pi/3$, when the cell becomes flat), and (ii) the rib length $b_\theta$, since all ribs are of the same length. The rib length can also be found from the face angle and a face diagonal size. In the case of the deformation considered, it is more convenient to use the diagonal length, which is the bisector of the angle $\theta$ where the wire lies. The length of that diagonal (wire) is \textit{the deformation invariant} and is equal to $a$.

For example, the unit cell of the \textit{original} interlaced wire metamaterial (the meshes are cubic) in Fig. \ref{fig:struct}(a) is described by the \textit{deformation angle} equal to $\acos(-1/3)$ (we will further use $\theta_\perp$ for that angle value), whereas the cubic unit cell in Fig. \ref{fig:unitcell}(b) appears to be described by the deformation angle of $\pi/2$.

%________________________________________Fig 3
%________________________________________

Thus, $\theta$ is the only necessary parameter to restore the structure geometry and will define the \textit{degree of deformation} (we will use below the terms \textit{deformation angle} for $\theta$ and $\theta$\textit{-material} (or \textit{-cell}) for the corresponding metamaterial configuration).

We describe analytically the deformation using a cubic cell as the initial one (Fig. \ref{fig:unitcell}(b)). Wherever we may have to refer to the original cell (Fig. \ref{fig:unitcell}(c)) as to the initial one, this will be specially indicated.
% In some places of the Article, we will use the description of the deformation, in which the original cell is taken as the initial one, but we will indicate it separately.

We compose a matrix $V_\theta$ of coordinates (in axes: $OX\parallel B_0A_0$ and $OY\parallel B_1B_2\parallel A_1A_2$, see Fig. \ref{fig:unitcell}) of \textit{$\theta$-cell} direct lattice vectors ($\boldsymbol{B_0A_j}(\theta)$, where $j=1,2,3$) for the arbitrary \textit{deformation angle} $\theta$: 
\begin{align}
    V_\theta=& \sum\limits_{j}\boldsymbol{B_0A_j}(\theta)\otimes\boldsymbol{\hat n_j},
	\label{eq:v_theta}
\end{align}
where $\boldsymbol{\hat n_j}$ are axes unit vectors ($j=1,2,3$)
Thus, our deformation can be represented as:
\begin{equation}
	V_\theta=\mathcal{D}_\theta\cdot V_{\pi/2},
\label{eq:dir_lat}
\end{equation}
where $\mathcal{D}_\theta$ is the operator that transform the coordinates matrix
\begin{equation}
    V_{\pi/2}=
	\left(
	\begin{array}{ccc}
		\sqrt{2} & \sqrt{2} & \sqrt{2} \\
		-\sqrt{3} & \sqrt{3} & 0 \\
		-1 & -1 & 2
	\end{array}\right) \frac{a}{2\sqrt{3}}
	\label{eq:v_halfpi}
\end{equation}
for the cubic-cell metamaterial (Fig. \ref{fig:unitcell}(b)) into the matrix $V_\theta$ for the \textit{$\theta$-material}.

$\mathcal{D}_\theta$ appears to be a diagonal matrix (deformation is an \textit{affine} transformation) in the chosen coordinate system and to depend on $\theta$ as:
\begin{equation}
	\mathcal{D}_\theta	= \operatorname{diag}\left(\frac{\sqrt{2\cos\theta+1}}{\sqrt{2} \cos \left(\theta / 2\right)}, \tan \left(\theta / 2\right), \tan \left(\theta / 2\right)\right).
\label{eq:dir_deformation}
\end{equation}

In Eq. (\ref{eq:dir_lat}) $V_\theta$ can be expressed by another \textit{initial} matrix $V_{\beta}$ (normally specified by an arbitrary angle $\beta$) in the right hand side of the equation. The idea is to choose a material with the known dispersion properties to further extrapolate them to the whole \textit{family}. The convenient choice will be $V_{\theta_\perp}=\mathcal{D}_{\theta_\perp}\cdot V_{\pi/2}$:
\begin{equation}
	V_\theta=
	\mathcal{D}_\theta V_{\pi/2}
	=
	\mathcal{D}_\theta
	\left(\mathcal{D}_{\theta_\perp}^{-1} V_{\theta_\perp}	\right) =
	\mathcal{I}_\theta V_{\theta_\perp},
	\label{eq:dir_lat_2}
\end{equation}
which implies that it is the unit cell of the \textit{original metamaterial} (the one with dispersion properties described in \cite{silveirinha2017lighttun, sakhno2021longitudinal}, Fig. \ref{fig:struct}(a)) that will be the \textit{initial} cell for the deformation. In this case the deformation matrix $\mathcal{I}_\theta$ in Eq. (\ref{eq:dir_lat_2}) is defined as:
\begin{equation}
	\mathcal{I}_\theta= \operatorname{diag}\left(\frac{\sqrt{2\cos\theta+1}}{\cos \left(\theta / 2\right)}, \frac{\tan \left(\theta / 2\right) }{\sqrt{2}}, \frac{\tan \left(\theta / 2\right) }{\sqrt{2}}\right)
	\label{eq:dir_lat_2_2}
\end{equation}
and provides the matrix of lattice vectors $V_\theta$ of $\theta$-metamaterial after applying it to the matrix $V_{\theta_\perp}$ for the \textit{original material}. 

Besides the \textit{deformation angle}, we introduce the following parameter which determines the \textit{degree of the deformation} of the \textit{original} metamaterial:
\begin{equation}
    p_{\mathrm{orig}}(\theta)=\frac{A_0B_0(\theta)}{A_0B_0(\theta_\perp)}=\mathcal{I}_\theta^{[1,1]}=\frac{\sqrt{2\cos\theta +1}}{\cos(\theta/2)},
    \label{eq:p_param}
\end{equation}
where $A_0B_0(\theta)$ is the length of the $\theta$-defined unit cell diagonal $A_0B_0$ (Fig. \ref{fig:unitcell}). Thus, for the \textit{original} geometry (Fig. \ref{fig:unitcell}(c), $\theta_3=\theta_\perp$) this parameter is equal to $1$, for the metamaterial with the cubic unit cell (Fig. \ref{fig:unitcell}(b), $\theta_2=\pi/2$) it equals $\sqrt{2}$.
%________________________________________

%________________________________________Low frequency points
\section{\label{sec:lfp} Low-frequency points}
Every material from the deformation \textit{family} consists of two non-connected identical wire meshes, therefore the low-frequency mode, which is supported by the \textit{original interlaced wire metamaterial}, exists for each family member. This can be shown using the quasistatic limit approach discussed in \cite{chen2018kpoints, sakhno2021longitudinal}.

%________________________________________Fig 4
\begin{figure*}
    \begin{minipage}{1\linewidth}
		\center{\includegraphics[width=1\textwidth]{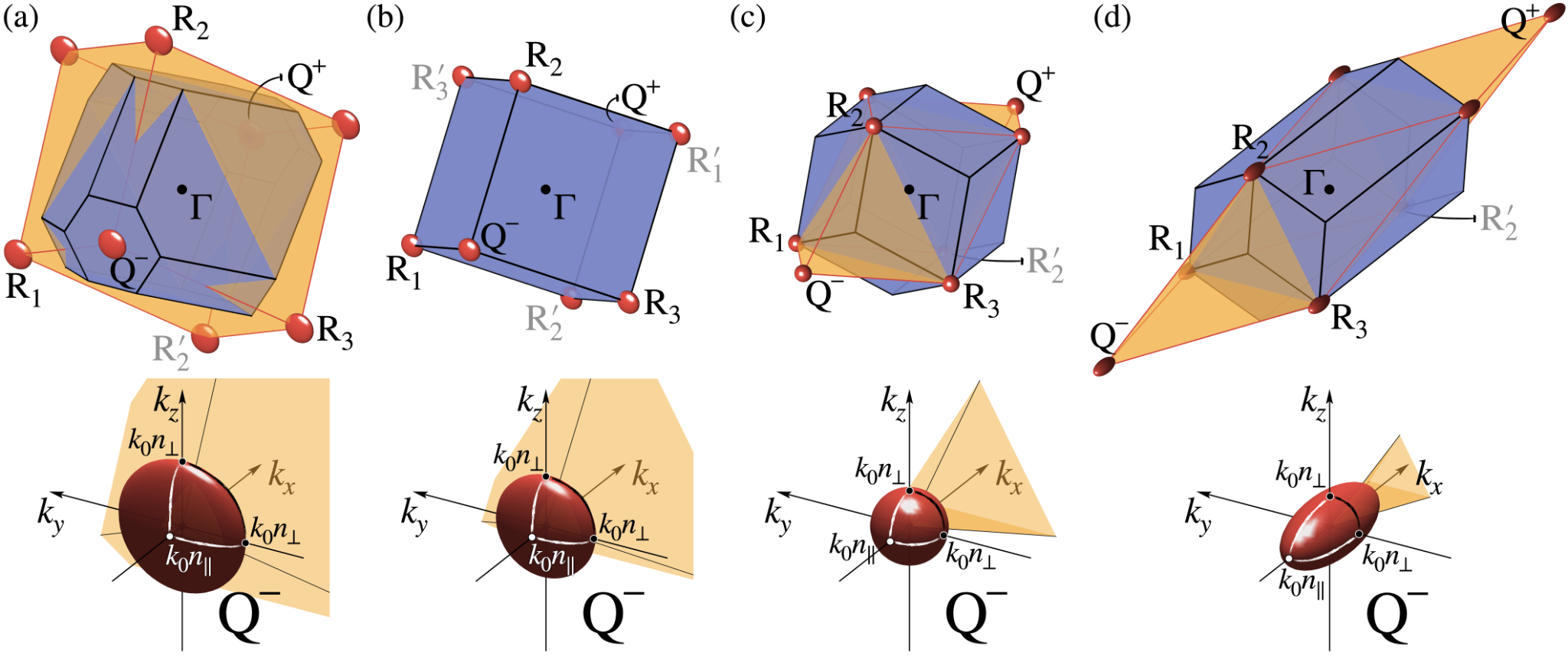} }
	\end{minipage}
	\caption{Brillouin zone (purple polyhedron), low frequency points location and shape of isofrequency surfaces for unit cell geometry with (a) $\theta_1 = 5\pi/12$; (b) $\theta_2 = \pi/2$; (c) $\theta_3 = \theta_\perp=\acos{\left(-1/3\right)}$; (d) $\theta_4 = 2\pi/3-\pi/90$.}
	\label{fig:b-zones}
\end{figure*}
%________________________________________

To define the reciprocal space points coordinates where zero-frequency mode appears, we refer to the unit cell geometry. In the quasistatic limit each mesh has its own potential ($\varphi_A$ and $\varphi_B$) since there is no electric contact between them. Bloch's theorem can be applied to the potential difference of meshes in each cell face, thus linking the differences on the opposite faces via the corresponding exponent (for example $(\varphi_B-\varphi_A)$ in $B_0A_2B_1A_3$ and $(\varphi_A-\varphi_B)$ in $BA_1B_3A_0B_2$ are linked by $\exp\left[i(\mathbf{k}_\theta\cdot \boldsymbol{B_0A_1}(\theta))\right]$, see Fig. \ref{fig:unitcell}). Thereby, the following system of equations can be composed:
\begin{equation}
	\begin{cases}
	(\varphi_B-\varphi_A)=(\varphi_A-\varphi_B) e^{i\left(\mathbf{k}_\theta\cdot \boldsymbol{B_0A_1}(\theta)\right)}\\
	(\varphi_B-\varphi_A)=(\varphi_A-\varphi_B) e^{i\left(\mathbf{k}_\theta\cdot \boldsymbol{B_0A_2}(\theta)\right)}\\
	(\varphi_B-\varphi_A)=(\varphi_A-\varphi_B) e^{i\left(\mathbf{k}_\theta\cdot \boldsymbol{B_0A_3}(\theta)\right)}
	\end{cases}.
	\label{eq:bloch_th}
\end{equation} 
Thus, the wave vectors $\mathbf{k}_\theta=(k_x,k_y,k_z)^\mathrm{T}(\theta)$ -- coordinates of low-frequency points for $\theta$-material -- that satisfy Eq. (\ref{eq:bloch_th}) can be defined by the following exhaustive condition:
\begin{equation}
    V_{\theta}^{\mathrm{T}}
	\mathbf{k}_\theta
	=\left(\mathcal{D}_\theta V_{\pi/2}\right)^{\mathrm{T}}
	\mathbf{k}_\theta
	=
	\begin{pmatrix}
		2n_1+1\\ 2n_2+1\\ 2n_3+1
	\end{pmatrix}
	\pi,
	\label{eq:bloch_th_2}
\end{equation}
where the superscript $\mathrm{T}$ stands for the transpose matrix.

Taking into account the diagonal form of the matrix $\mathcal{D}_{\theta}$ (Eq. (\ref{eq:dir_deformation})) the solution of Eq. (\ref{eq:bloch_th_2}) can be expressed as:
\begin{equation}
	\mathbf{k}_{\theta}=
	\mathcal{D}_{\theta}^{-1} \mathbf{k}_{\pi/2},
\label{eq:lfp_location_0-5pi}
\end{equation}
where $\mathbf{k}_{\pi/2}$ are the coordinates of low-frequency points in $k$-space for the metamaterial with the cubic unit cell, i.e. the solution of Eq. (\ref{eq:bloch_th_2}) with the substitution of the unit matrix $\mathcal{D}_{\pi / 2}$ and the matrix $V_{\pi / 2}$.

We can also express $\mathbf{k}_{\theta}$ in terms of the $\mathbf{k}_{\theta_\perp}$ coordinates for the original interlaced wire metamaterial ($\theta_\perp$-material, Fig. \ref{fig:unitcell}(c)) and rewrite Eq. (\Ref{eq:lfp_location_0-5pi}) as:
\begin{equation}
	\mathbf{k}_{\theta}=
	\mathcal{D}_{\theta}^{-1} \mathbf{k}_{\pi/2}=
	\mathcal{D}_{\theta}^{-1} \left(\mathcal{D}_{\theta_\perp} \mathbf{k}_{\theta_\perp}\right)=\mathcal{K}_{\theta} \mathbf{k}_{\theta_\perp}.
\label{eq:lfp_location_iwmm}
\end{equation}
This representation will allow us to further refer to the shape of isofrequency surfaces, since for the original metamaterial isofrequency the surfaces have spherical shapes at low frequencies \cite{sakhno2021longitudinal}.
%________________________________________

%________________________________________Low-frequency points coordinates
\section{ Coordinates of low-frequency points} \label{sec:lfp-movement}
\begin{figure*}
    \begin{minipage}{1\linewidth}
        \center{\includegraphics[width=1\textwidth]{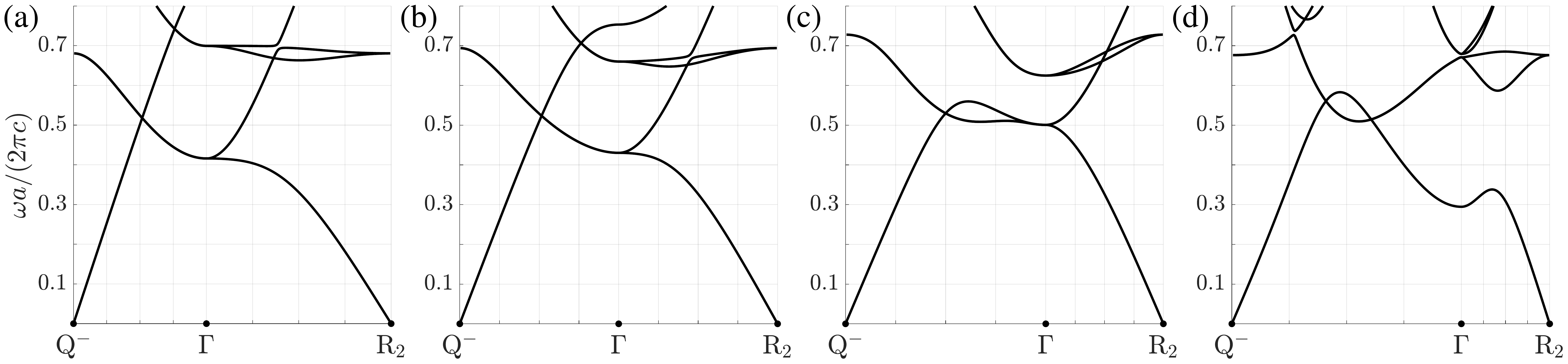} }
    \end{minipage}
    \caption{Dispersion $\mathrm{Q}\Gamma\mathrm{R}$-path (see Fig. \ref{fig:b-zones}) diagrams for the metamaterial defined by \textit{the deformation angle}: (a) $\theta_1=5\pi/12$; (b) $\theta_2=\pi/2$; (c) $\theta_3=\theta_\perp=\operatorname{acos}(-1/3)$; (d) $\theta_4=2\pi/3-\pi/90$.}
    \label{fig:disp-qgr}
\end{figure*}

The relative position of the Brillouin zone of the deformed metamaterial and its low-frequency points can be visualized in $k$-space (see Fig. \Ref{fig:b-zones}). The absence of the $\Gamma$-born surfaces for the metamaterials of the considered \textit{family} is confirmed by Fig. \Ref{fig:b-zones} and also by the computed (using thin PEC wires in COMSOL\cite{comsol} model) dispersion graphs presented in Fig. \ref{fig:disp-qgr}.

For the cubic unit cell metamaterial ($\theta_2 = \pi / 2$, Fig. \ref{fig:unitcell}(b)) the Brillouin zone is a cube (\textit{Simple Cubic Lattice}), and isofrequency surfaces are born in its eight corner points (see Fig. \ref{fig:b-zones}(b)): two of them in $\mathrm{Q}$-points (they lie on the $k_x$ axis, symmetrically with respect to the point $\Gamma$), while the other six are at $\mathrm{R}$-points (outside the $k_x$ axis). 

The slightest change in the deformation angle $\theta$ results in a different relative position of the $k$-spheroids and the Wigner-Seitz cell shape in the reciprocal space (blue polyhedrons in Fig. \ref{fig:b-zones}). 

%________________________________________Fig 5
%________________________________________

For $\theta <\pi / 2$ (Fig. \ref{fig:unitcell}(a)) $\mathrm{Q}$-points lie on the surface of the Brillouin zone of the \textit{RHL1} type (\textit{Rhombohedral Lattice type 1}, Fig. \ref{fig:b-zones}(a)), and $\mathrm{R}$-ellipsoids approach its corners with the deformation angle $\theta$ approaching $\pi / 2$. 

For angles $\theta>\pi / 2$ the opposite is true: $\mathrm{R}$-ellipsoids are located in the corners of the \textit{RHL2} cell (in the case of $\theta = \theta_\perp$ there is a body-centered cubic Bravais lattice; the Brillouin zone has the \textit{BCC} type, Fig. \ref{fig:b-zones}(c)), and $\mathrm{Q}$-ellipsoids are moving away from its vertices with the deformation angle increasing (Fig. \Ref{fig:b-zones}(c,d)).

In the chosen coordinate system, the $\mathrm{Q}$-ellipsoids birth points upon deformation of the cubic cell will only move along the $k_x$ axis, and their coordinates can be described by the first non-zero element of the diagonal matrix, inverse to the matrix $\mathcal{D}_\theta$ according to Eq. (\ref{eq:lfp_location_0-5pi}):
\begin{equation}
	\boldsymbol{\Gamma\mathrm{Q}}^{\pm}(\theta) =
	\mathcal{D}_{\theta}^{-1} \boldsymbol{\Gamma\mathrm{Q}}^{\pm}\left(\pi/2\right)=
	\begin{pmatrix}
	\pm \frac{\sqrt{3} \cos \left(\theta / 2\right)}{\sqrt{2\cos\theta+1}}\\ 0\\	0
	\end{pmatrix}\frac{2\pi}{a}.
	\label{eq:x_pnt_mvmnt}
\end{equation}
The description of the $\mathrm{R}$-points evolution in accordance with the same Eq. (\ref{eq:lfp_location_0-5pi}) will be slightly more complicated, since the coordinates of these points in the $k$-space do not seem to be trivially defined:
\begin{align}
	\boldsymbol{\Gamma\mathrm{R}_j}(\theta) &=
	\mathcal{D}_{\theta}^{-1} \boldsymbol{\Gamma\mathrm{R}_j}(\pi/2)
 	\nonumber \\
	&=\mathcal{D}_{\theta}^{-1} \left[
	\boldsymbol{\Gamma\mathrm{Q}}^{\pm}(\pi/2) \mp \boldsymbol{\mathrm{Q}\mathrm{R}_j}(\pi/2) 
	\right],
	\label{eq:h_pnt_mvmnt}
\end{align}
where $j=1,2,3$, and the reciprocal vectors $\boldsymbol{\mathrm{Q}\mathrm{R}_j}(\pi/2)$ have the following coordinates (see Fig. \ref{fig:b-zones}(b)):
\begin{align}
    & \boldsymbol{\mathrm{Q}\mathrm{R}_1}(\pi/2) = \frac{1}{\sqrt{3}}
	\left(\begin{array}{ccc}
		\sqrt{2}\\ \sqrt{3}\\ -1
	\end{array}\right) \frac{2\pi}{a}, \nonumber \\
	& \boldsymbol{\mathrm{Q}\mathrm{R}_2}(\pi/2) = \frac{1}{\sqrt{3}}
	\left(\begin{array}{ccc}
		\sqrt{2}\\ 0\\ 2
	\end{array}\right) \frac{2\pi}{a}, \nonumber \\
	& \boldsymbol{\mathrm{Q}\mathrm{R}_3}(\pi/2) = \frac{1}{\sqrt{3}}
	\left(\begin{array}{ccc}
		\sqrt{2}\\ -\sqrt{3}\\ -1
	\end{array}\right) \frac{2\pi}{a}.
\end{align}

%________________________________________Fig 6
\begin{figure}[h!]
    \center{\includegraphics[width=1\linewidth]{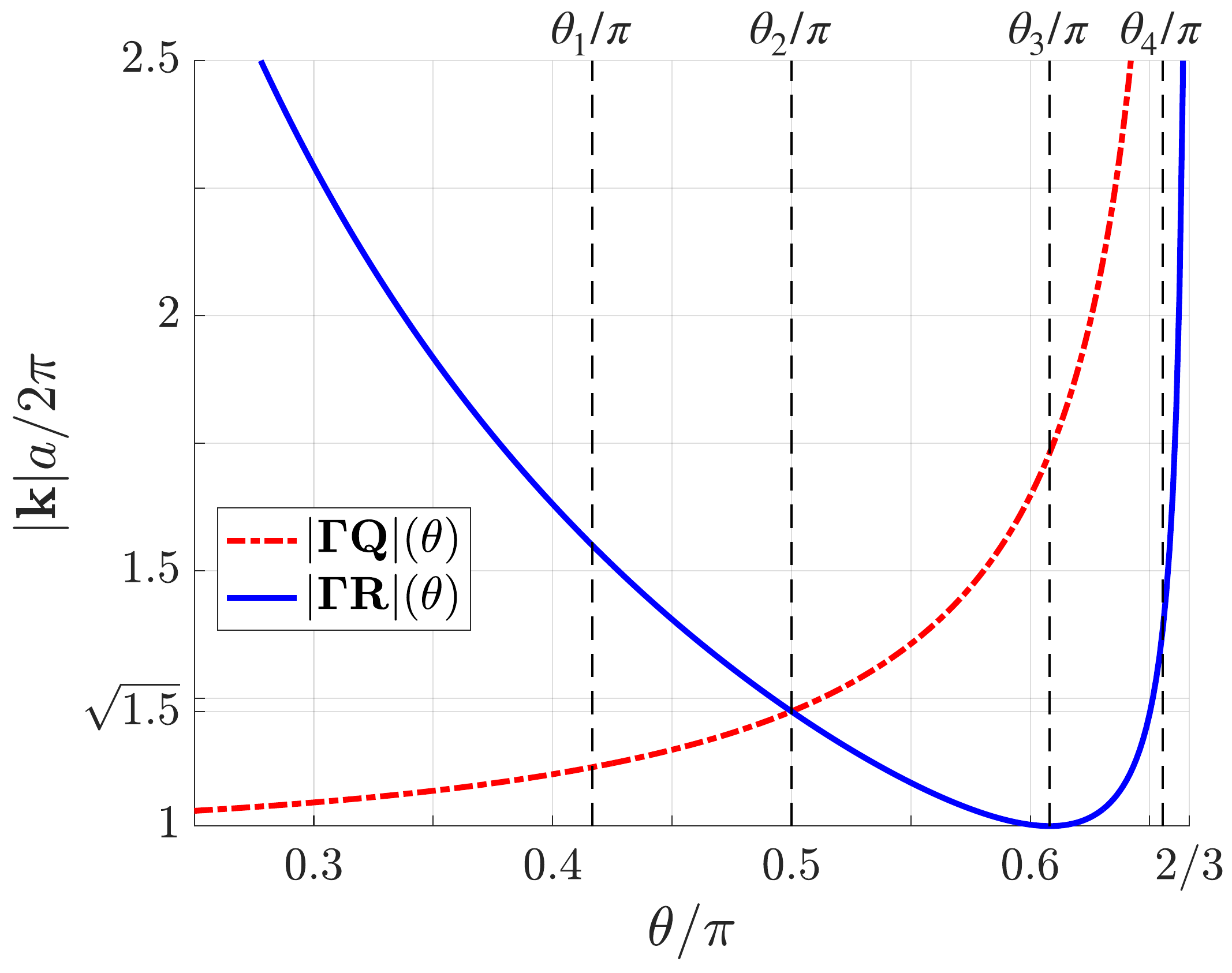} }
    \caption{ The distance of low-frequency points from $\Gamma$ point plotted versus $\theta$. Dash-dotted red line describes the dependence of $|\boldsymbol{\Gamma\mathrm{Q}}|(\theta)$ (see Eq. (\ref{eq:x_pnt_mvmnt})) on deformation angle, the solid blue line that of $|\boldsymbol{\Gamma\mathrm{R}}|(\theta)$ (see Eq. \ref{eq:h_pnt_mvmnt})). Vertical lines show some geometries of the unit cell from Fig. \ref{fig:unitcell} (black dashed lines).	}
    \label{fig:lfp-k-vs-theta}
\end{figure}
%________________________________________

Dependencies of $\mathrm{Q}$ and $\mathrm{R}$ low frequency points distances from $\Gamma$ point ($|\boldsymbol{\Gamma\mathrm{Q}}^\pm|(\theta)$ and $|\boldsymbol{\Gamma\mathrm{R}}|(\theta)$ from Eqs. (\ref{eq:x_pnt_mvmnt}) and (\ref{eq:h_pnt_mvmnt})) on the deformation angle $\theta$ are plotted in Fig. \Ref{fig:lfp-k-vs-theta} by solid red and dashed blue lines respectively.

%________________________________________Table 1
\begin{table}[h!]
    \centering
    \begin{tabular}{|c|c|c|c|}
        \hline
         $\theta$ &  $p_\text{orig}$ & $g_{_\text{\text{BZ}}}$ & Brillouin zone type\\
         \hline\hline
         $\theta_1$ & 1.553 & 0.72 & RHL1\\
         $\theta_2=\pi/2$ & $\sqrt{2}$ & 1 & SCL\\
         $\theta_3=\theta_\perp$ & 1 & 1.732 & BCC\\
         $\theta_4$ & 0.480 & 2.6 & RHL2\\
         \hline
    \end{tabular}
    \caption{Table of $p_\text{orig}$ (Eq. \ref{eq:p_param}) and $g_{_\text{BZ}}$ (Eq. \ref{eq:g_param}) parameters for different $\theta$-materials.}
    \label{tab:params}
\end{table}
%________________________________________

Next, we introduce a parameter in the reciprocal space, which characterises the ratio of the distance from $\Gamma$-point to the nearest low-frequency point located along $x$-axis, the $\mathrm{Q}$-point, to the distance to the nearest out-of axis isofrequency surface center -- one of the $\mathrm{R}_j$ points (where $j=1,2,3$). This parameter will indirectly condition the Brillouin zone geometry:
\begin{equation}
    g_{_{\text{BZ}}}(\theta)=\frac{|\boldsymbol{\Gamma\mathrm{Q}}(\theta)|}{|\boldsymbol{\Gamma\mathrm{R}_j}(\theta)|}.
    \label{eq:g_param}
\end{equation}
%________________________________________Fig 7
\begin{figure*}
    \center{\includegraphics[width=1\textwidth]{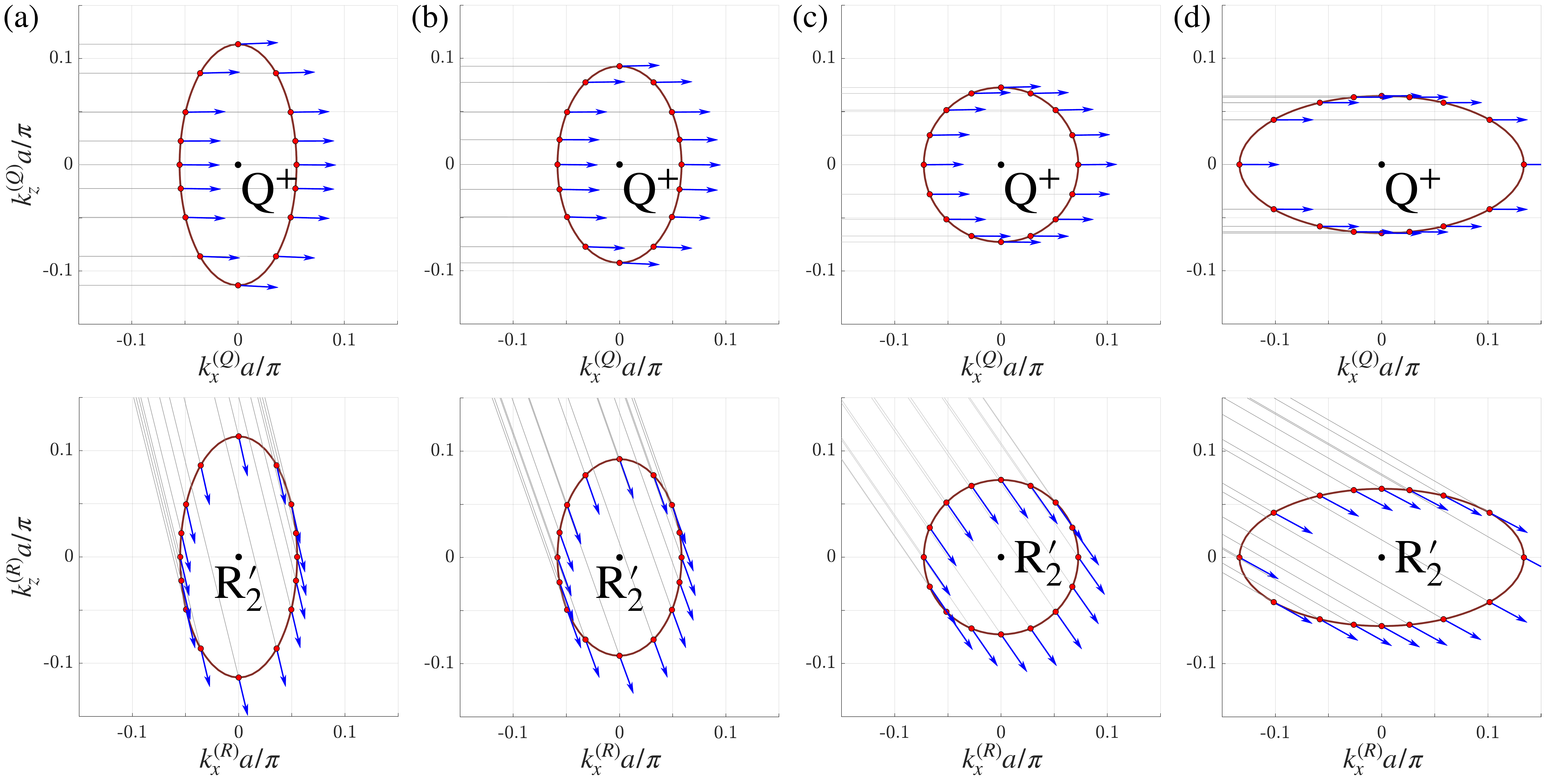}}
    \caption{Isofrequency contours ($\omega a/(2\pi c)=0.025$) around $Q^+$ and $\mathrm{R}_2'$ points of the reciprocal space (Fig. \ref{fig:b-zones}) with average electric field $\mathbf{E}^{av}$ shown (blue arrows) for unit cell geometry ($a=1$cm) with (a) $\theta_1 = 5\pi/12$; (b) $\theta_2 = \pi/2$; (c) $\theta_3 = \theta_\perp=\acos{\left(-1/3\right)}$; (d) $\theta_4 = 2\pi/3-\pi/90$. Solid grey lines stand for the wave vectors direction (from $\Gamma$-point to the corresponding contour point).}
    \label{fig:isofreq-cont-av}
\end{figure*}
%________________________________________
For example, the cubic Brillouin zone (Fig. \ref{fig:b-zones}(b)) corresponds to the metamaterial with the cubic unit cell (Fig. \ref{fig:unitcell}(b)) and $g_{_{\text{BZ}}}(\pi/2)=1$. The parameter becomes less then $1$ after any stretching of the cubic-cell metamaterial, i.e. for any metamaterial from the \textit{family} with $\theta<\pi/2$. It is greater than $1$ for materials with $\theta>\pi/2$ (see Table \ref{tab:params}).

Analytical forecasting of the low-frequency points location during the unit cell deformation makes it possible to consider a zone in the $k$-space bounded by a hexahedron whose vertices are the two $\mathrm{Q}$-points and the six $\mathrm{R}$-points. This polyhedron is highlighted in yellow in Fig. \Ref{fig:b-zones}: for any value of the angle $\theta$ its volume is equal to the volume of the Wigner-Seitz cell in reciprocal space. For $\theta = \pi / 2$, the polyhedron corresponds to the Wigner-Seitz cell, i.e. the polyhedron in the case of any $\theta$ is essentially the deformation $\mathcal{D}_{\theta}^{-1}$ applied to the Wigner-Seitz cell for the cubic unit cell. So this polyhedron is the Brillouin zone, the coordinates and shape of which are described easier (by Eqs. (\ref{eq:x_pnt_mvmnt}) and (\ref{eq:h_pnt_mvmnt})) than the coordinates and shape of the Wigner-Seitz cell for any \textit{deformation angle}.
%________________________________________

%________________________________________Isofrequency surfaces and polarization
\section{\label{sec:lfs-shape} Isofrequency surfaces and polarization}	
The spherical form of isofrequency surfaces confirmed for the original interlaced wire metamaterial ($\theta=\theta_\perp$) at low frequencies \cite{sakhno2021longitudinal} allows to predict the shape of isofrequency surfaces for an arbitrary deformed configuration. % in $k$-space 

The deformation matrix $\mathcal{K}_{\theta}$ from Eq. (\ref{eq:lfp_location_iwmm}) has the form:
\begin{align}
	\mathcal{K}_\theta=
	\operatorname{diag}\left(
	\frac{\cos \left(\theta / 2\right)}{\sqrt{2\cos\theta+1}}, \sqrt{2}\cot \left(\theta / 2\right),
	\sqrt{2}\cot \left(\theta / 2\right)\right)
\label{eq:iwmm_k_deformation}
\end{align}
and is the inverse of the matrix $\mathcal{I}_\theta$ from Eq. (\ref{eq:dir_lat_2}). Applying it to a sphere in $k$-space, we obtain a spheroid (oblate or prolate depending on $\theta$) with $k_x$ as the symmetry axis (Fig. \Ref{fig:b-zones}). 
Thus, we can introduce the longitudinal $n_\parallel$ and transverse $n_\perp$ refractive indices at low frequencies from the shape of the isofrequency surfaces:
\begin{equation}
	\left|\mathbf{k}_\parallel\right|=
	\left|k_x\right|=k_0n_\parallel,\quad
	\left|\mathbf{k}_\perp\right|=
	\sqrt{k_y^2+k_z^2}=k_0 n_\perp,
	\label{eq:n_def}
\end{equation}
where $|\mathbf{k_\parallel}|$ is the distance from the centre to the pole along the symmetry axis, $|\mathbf{k_\perp}|$ is the equatorial radius of the spheroid, $k_0$ is the vacuum wavenumber.

\begin{figure}[h!]
    \center{\includegraphics[width=1\linewidth]{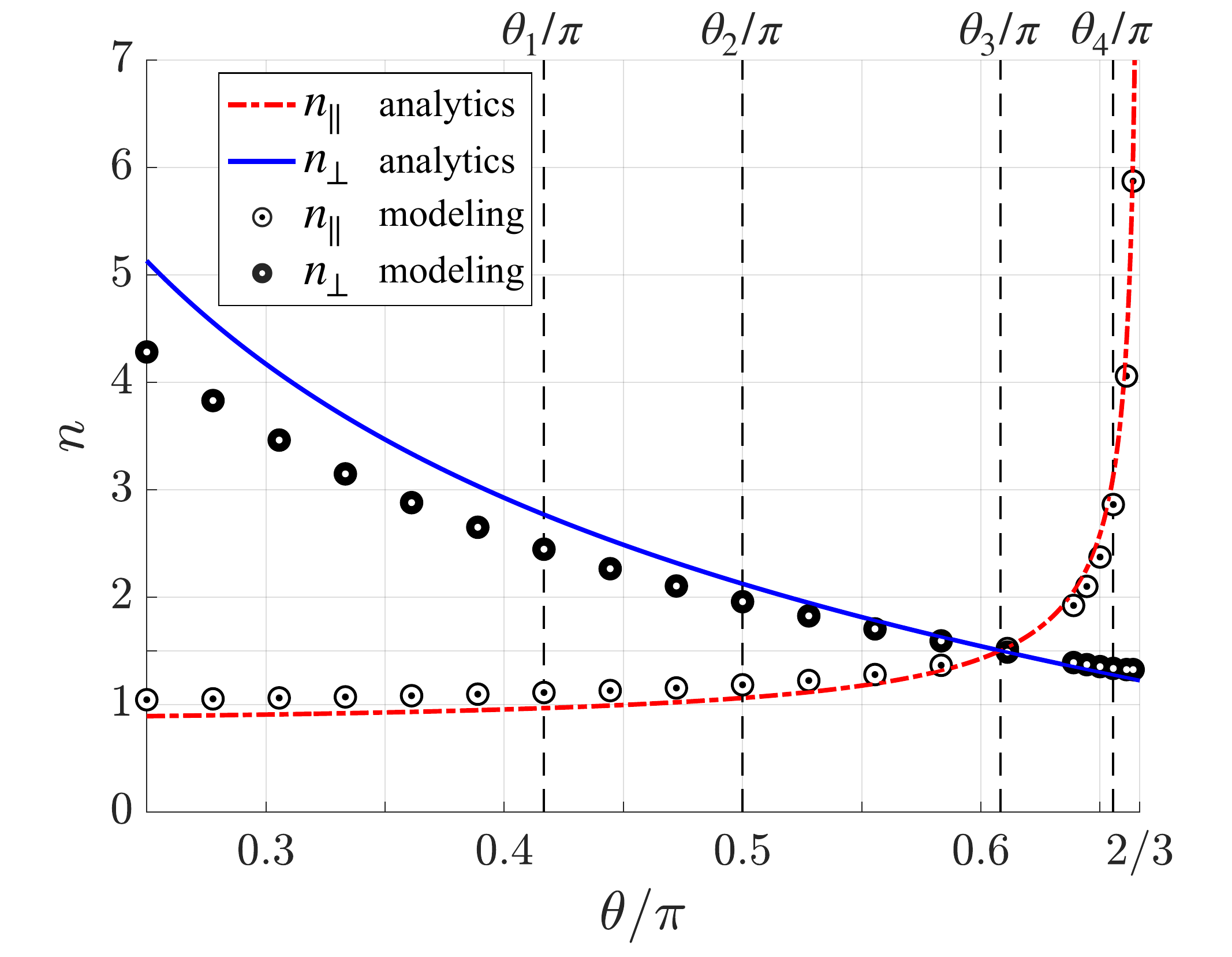} }
    \caption{Refractive indices versus $\theta$ plot: (1) $n_\parallel$ (red dash-dotted line) and $n_\perp$ (solid blue line) obtained analytically when $\mathcal{I}_\theta$ from Eq. (\ref{eq:dir_lat_2_2}) was applied to unit cell geometry and the background medium (using \textit{transformation optics} theory, Eq. (\ref{eq:to_vacuum})); (2) $n_\parallel$ (white circles) and $n_\perp$ (black circles) obtained from the simulation when deformation was applied to unit cell geometry only (vacuum is undeformed).}
    \label{fig:n-theta-plot}
\end{figure}

We performed numerical simulations in COMSOL 5.5 of the structure with thin wires (of the length $a=1$cm) from PEC, where the deformation operator $\mathcal{D}_\theta$ acted only on the geometry of the unit cell. Thus, the dependence of the indices $n_\parallel$ and $n_\perp$ values on $\theta$ can be revealed and plotted as a graph in Fig. \ref{fig:n-theta-plot} (black and white circles). The physical properties of the medium (vacuum) where the metamaterial is located did not change during these calculations and the environment remained isotropic, i.e. the transformation optics theory \cite{pendry2006controlling,sun2017transformation}  has not been applied. Hence, there appears a noticeable discrepancy ($15-20\%$) between the simulation results and the theoretical dependencies of the corresponding matrix $\mathcal{K}_{\theta}$ terms (red dash-dotted and blue solid lines in Fig. \ref{fig:n-theta-plot}). 

%________________________________________Fig 8
%________________________________________

The $\mathcal{K}_{\theta}$ matrix will accurately predict the shape of the isofrequency surfaces upon deformation of the original cell (in this case -- the original interlaced wire metamaterial cell) only when the host medium parameters $\overset{=}{\eta}_{host}$ (tensor $\overset{=}{\varepsilon}_{host}$ or $\overset{=}{\mu}_{host}$) will change according to the transformation optics theory \cite{pendry2006controlling, sun2017transformation}:
\begin{equation}
	\overset{=}{\varepsilon}_{\text{host}}(\theta)=
	\overset{=}{\mu}_{\text{host}}(\theta)=
	\frac{\mathcal{I}_\theta \mathds{1} \mathcal{I}_\theta^{T}}{\operatorname{det} \mathcal{I}_\theta},
	\label{eq:to_vacuum}
\end{equation}
where $\mathds{1}$ is  a unit tensor which is standing for $\overset{=}{\varepsilon}_{\text{vac}}$ (or $\overset{=}{\mu}_{\text{vac}}$) -- permittivity (or permeability) tensors of the vacuum. If such \textit{vacuum deformation} is taken into account in simulations, similar dependencies will meet the theoretical estimate. 

%\section{\label{sec:conclusion} Polarization}
The final part of our research is field polarization verification for the low-frequency mode supported by the deformed material. As mentioned in the Introduction, the corresponding results for \textit{the original interlaced wire metamaterial}, which confirm the longitudinal polarization of the mode, were presented in our paper \cite{sakhno2021longitudinal} that also provides a methodology for such calculations.

Isofrequency contours for various deformed configurations in $\Gamma\mathrm{Q}^+\mathrm{R}_2$ plane of the reciprocal space (in the vicinity of $\mathrm{Q}^+$ and $\mathrm{R}_2'$ points, see Fig. \ref{fig:b-zones}) are shown in Fig. \ref{fig:isofreq-cont-av}. The original isofrequency circle (Fig. \ref{fig:isofreq-cont-av} (c)) does not split into two curves during the metamaterial deformation, due to its longitudinal polarization.

Contours shape (Fig. \ref{fig:isofreq-cont-av}) remain ellipses that shrink (or stretch) along the $k_x$-axis during the structure deformation, and the longitudinal polarisation of the mode is retained (which indirectly confirms its scalar nature \cite{fan2007nonmax}). 
%________________________________________

%________________________________________ Discussion
\section{Discussion}
Spherical isofrequency surfaces are typical for isotropic materials. Although with the original wire metamaterial surfaces do not originate at $\Gamma$-point, as was shown above (Eq. (\ref{eq:lfp_location_0-5pi}), Fig. \ref{fig:b-zones}), we observe similarity between the original configuration of the material and the artificial environment of metal spheres located at the nodes of the body-centered cubic lattice.

Since the polarizabilities for a sphere are the same in all directions, the BCC metamaterial from the spheres will be isotropic (the mode with a spherical isofrequency surface is born at $\Gamma$-point), and the mode itself will degenerate in polarization due to the identity of the magnetic and electric polarizabilities (see ch. 12 in \cite{collin1990field}).

Under the deformation (Eq. \ref{eq:dir_lat_2_2}) the BCC symmetry of the lattice will be violated and prolate/oblate spheroids will be located at the nodes of the deformed lattice. For a spheroid, the statement about the equivalence of directions becomes incorrect, since the polarizability along stretching/compression axis will differ from the polarizabilities in other directions \cite{collin1990field}. Magnetic and electrical polarizabilities are no longer equal to each other. Thus, the mode stops degenerating and splits: the material acquires the pre-determined direction (the optical axis) along the deformation axis.
%________________________________________

%________________________________________Conclusion
\section{\label{sec:conclusion} Conclusion}
In this study, we considered the deformations of a metamaterial consisting of two independent metal meshes. We showed that the type of deformations described results in a theoretically predictable shift of the points where isofrequency surfaces are born. Isofrequency surfaces have an analytically predictable shape of ellipsoids, the dimensions of their semi-axes changing monotonically with a change in the deformation angle $\theta$ and becoming identical for the original interlaced wire metamaterial. On the other hand, the low-frequency mode polarisation stays unchanged during the deformation and remains longitudinal.
%________________________________________

\section*{Acknowledgments}
The authors acknowledge Professor Maxim Gorlach for his help and productive discussions. This work was supported by the Ministry of Science and Higher Education of the Russian Federation (Project No. 075-15-2022-1120). E.K. acknowledges the RPMA grant from the School of Physics and Engineering of ITMO University.

%\appendix 
%\input{appendix}

\newpage

\bibliographystyle{ieeetr}
\bibliography{bib}
\end{document}